\documentclass[aps,prl,twocolumn,superscriptaddress]{revtex4-1}
\usepackage{graphicx}
\usepackage{color}
\usepackage{amsfonts}
\usepackage{isomath}
\usepackage{amsmath}
\usepackage{amsthm}
\usepackage{amsfonts}
\usepackage{braket}

\usepackage{hyperref}

\usepackage{yhmath}
\usepackage{gensymb}

\def\ReSe2{ReSe$_2$}

\renewcommand{\vec}[1]{\mathbfit{#1}}

\begin{document}

\title{External screening and lifetime of exciton population in single-layer \ReSe2 probed by time- and angle-resolved photoemission spectroscopy}

\author{Klara Volckaert}
\affiliation{Department of Physics and Astronomy, Interdisciplinary Nanoscience Center, Aarhus University, 8000 Aarhus C, Denmark}
\author{Byoung Ki Choi}
\affiliation{Department of Physics, University of Seoul, Seoul 02504, Republic of Korea}
\affiliation{Advanced Light Source, E. O. Lawrence Berkeley National Laboratory, Berkeley, California 94720, USA}
\author{Hyuk Jin Kim}
\affiliation{Department of Physics, University of Seoul, Seoul 02504, Republic of Korea}
\author{Deepnarayan Biswas}
\affiliation{Department of Physics and Astronomy, Interdisciplinary Nanoscience Center, Aarhus University, 8000 Aarhus C, Denmark}
\author{Denny Puntel}
\affiliation{Dipartimento di Fisica, Università degli Studi di Trieste, 34127 Trieste, Italy}
\author{Simone Peli}
\affiliation{Elettra-Sincrotrone Trieste S.C.p.A., 34149 Basovizza, Trieste, Italy}
\author{Fulvio Parmigiani}
\affiliation{Dipartimento di Fisica, Università degli Studi di Trieste, 34127 Trieste, Italy}
\affiliation{Elettra-Sincrotrone Trieste S.C.p.A., 34149 Basovizza, Trieste, Italy}
\author{Federico Cilento}
\affiliation{Elettra-Sincrotrone Trieste S.C.p.A., 34149 Basovizza, Trieste, Italy}
\author{Young Jun Chang}
\affiliation{Department of Physics, University of Seoul, Seoul 02504, Republic of Korea}
\affiliation{Department of Smart Cities, University of Seoul, Seoul 02504, Republic of Korea}
\author{S{\o}ren Ulstrup}
\email{ulstrup@phys.au.dk}
\affiliation{Department of Physics and Astronomy, Interdisciplinary Nanoscience Center, Aarhus University, 8000 Aarhus C, Denmark}

\begin{abstract}
The semiconductor \ReSe2{} is characterized by a strongly anisotropic optical absorption and is therefore promising as an optically active component in two-dimensional heterostructures. However,  the underlying femtosecond dynamics of photoinduced excitations in such materials has not been sufficiently explored.  Here, we apply an infrared optical excitation to single-layer \ReSe2{} grown on a bilayer graphene substrate and monitor the temporal evolution of the excited state signal using time- and angle-resolved photoemission spectroscopy. We measure an optical gap of $(1.53 \pm 0.02)$ eV,  consistent with resonant excitation of the lowest exciton state.  The exciton distribution is tunable via the linear polarization of the pump pulse and exhibits a biexponential decay with time constants given by $\tau_1 = (110 \pm 10)$ fs and $\tau_2 = (650 \pm 70)$ fs, facilitated by recombination via an in-gap state that is pinned at the Fermi level.  By extracting the momentum-resolved exciton distribution we estimate its real-space radial extent to be greater than 17.1 \AA{},  implying significant exciton delocalization due to screening from the bilayer graphene substrate. 
\end{abstract}

\maketitle

Among the semiconducting transition metal dichalcogenides (TMDCs), rhenium-based compounds such as \ReSe2{} are distinctive due to their anisotropic 1T$^{\prime}$ structure, resulting from an in-plane Jahn-Teller-like distortion \cite{kertesz1984octahedral}. The lattice structure can be thought of as one-dimensional chains of Re atoms, reducing the three-fold symmetry of a common equilateral hexagonal lattice that describes most other semiconducting TMDCs \cite{manzeli20172d}.  The reduced symmetry introduces a linear anisotropy for optical absorption  \cite{Zhong2015}.  Consequently, \ReSe2{} stands out as a potential candidate material for multifunctional two-dimensional optoelectronic devices, such as photodetectors \cite{Zhang2016,Afzal:2021,Li:2022} and field effect transistors \cite{Yang:2014,Corbet2016,Pradhan:2018,Rehman:2021}. 

The electronic structure of \ReSe2{} exhibits a direct quasiparticle gap at the $\mathrm{\Gamma}$-point of the Brillouin zone (BZ),  which has been calculated to be 1.49 eV in the bulk and 2.44 eV in a free-standing single layer (SL) \cite{Arora2017}.  Scanning tunneling spectroscopy (STS) measurements of SL \ReSe2{} have shown a gap of 1.7 eV when the material is supported on bilayer graphene on silicon carbide \cite{Choi2020,Trinh:2022} and 2.0 eV when supported on graphene on hexagonal boron nitride \cite{Qiu2019}. This variation in gap sizes underscores the important role of substrate screening.  Photoluminescence measurements have revealed an optical gap of 1.39 eV in thick films that increases to 1.51 eV in SL \ReSe2{}, corresponding to the lowest exciton transition in the system \cite{Arora2017,Kipczak:2020,Wolverson:2014,Zhao2015}.  

The dynamics of photoinduced carriers has been investigated in multilayer and bulk \ReSe2{} via pump-probe measurements of the differential transmissivity at pump wavelengths of 400 and 800 nm, revealing biexponential decay dynamics on timescales of 10 ps and 80 ps caused by the interplay of exciton formation and decay with hot carrier generation and relaxation \cite{Liu:2019,He:18}.  In exfoliated SL \ReSe2{} on insulating subtrates the differential transmittivity exhibits dynamics on the order of 10 ps \cite{He:18}.  Integration of SL ReSe$_2$ with SL MoS$_2$ in a vertical heterostructure enables further tuning of photoinduced carrier lifetimes via the formation of  interlayer excitons with relaxation time constants exceeding 300 ps \cite{Yang:2021}.  

Further insights to the light-matter interaction in SL \ReSe2{} are required in order to design optoelectronic devices.  Indeed, the ultrafast dynamics and excitonic properties of SL \ReSe2{} in the presence of conductive graphene layers has not been studied. 

Here, we use time- and angle-resolved photoemission spectroscopy (TR-ARPES) in order to determine the ultrafast dynamics of excitons in SL \ReSe2{} epitaxially grown on electron-doped bilayer graphene. By applying an 800 nm optical excitation, we observe a significant population of both exciton and in-gap states, with a strong dependence of the exciton signal on the polarisation of the pump pulse.  The exciton population exhibits an ultrafast biexponential decay with time constants of $\tau_1 = (110 \pm 10)$ fs and $\tau_2 = (650 \pm 70)$ fs. Our momentum-resolved measurements of the exciton distribution lead to a lower estimate of the exciton radius of 17.1 \AA{} on bilayer graphene, indicating that the substrate screening causes a relatively high degree of exciton delocalization compared to bulk \ReSe2{}, where the radius is 9.6 \AA{} \cite{Arora2017}.

SL \ReSe2{} is grown using molecular beam expitaxy on a 6H-SiC substrate, which is annealed at 1573 K to form a bilayer of graphene via thermal decomposition of SiC. Re and Se atoms are co-evaporated from an e-beam evaporator and an effusion cell, respectively, at a sample temperature of 523 K for 10 min. Following annealing at 693 K for 30 min, a 0.8 ML \ReSe2{} layer is formed with three-fold rotated domains, as described in detail in Ref. \cite{Choi2020}.  The bilayer graphene is substantially electron-doped with a carrier concentration of 1.6 $\times$ 10$^{13}$ cm$^{-2}$ \cite{Choi2020}. The sample is encapsulated in a thick layer of Se to avoid contamination during transfer through air to the photoemission experiment.

\begin{figure}[t!] 
\begin{center}
\includegraphics[width=0.49\textwidth]{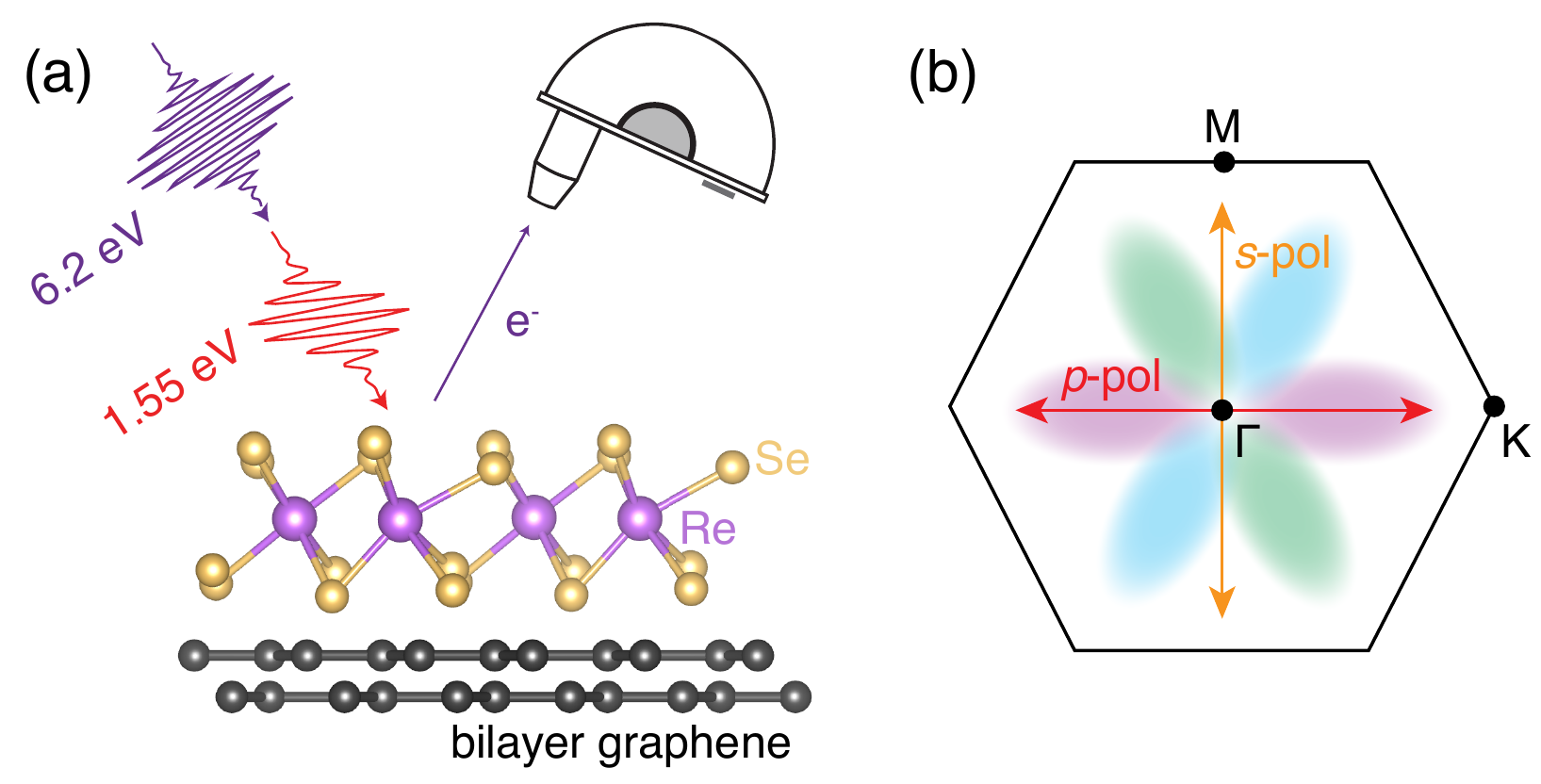}
\caption{(a) Sketch of TR-ARPES experiment on SL \ReSe2{} on bilayer graphene. (b) Diagram of SL \ReSe2{} BZ with definition of pump pulse polarisation in our geometry.  Our SL \ReSe2{} synthesis results in three rotated domains \cite{Choi2020},  with each additional domain contributing to the oscillator strength of the $\mathrm{\Gamma}$ exciton \cite{Zhong2015},  as shown via pink,  blue and green ovals.}
\label{fig:1}
\end{center}
\end{figure}

The TR-ARPES measurements are performed at the T-ReX facility in Trieste, Italy.  Equilibrium ARPES measurements of the valence band (VB) are performed using the ninth harmonic at 10.8 eV of a Yb fiber laser \cite{Peli2020}.  For time-resolved measurements, a 250 kHz Ti:sapphire laser with a fundamental energy of 1.55 eV is split in two time-delayed pulses with the fundamental beam applied as a pump pulse and the other part used as a probe pulse at 6.2~eV via generation of the fourth harmonic.  The probe pulses are always $p$-polarized. Photoemitted electrons are detected using a SPECS Phoibos 225 analyser.  An overview of the experiment is presented schematically in Fig. \ref{fig:1}(a). The time, energy and angular resolution are set to 150~fs,  50 meV and 0.2$^{\circ}$, respectively.  The Se capping layer on the \ReSe2{} sample is removed via one hour of annealing to a sample temperature of 473~K prior to TR-ARPES measurements.  The sample is kept at 120~K throughout the measurements unless stated otherwise. 

The linear polarisation of the pump pulse is varied between $s$ and $p$ via a half-wave plate.  \ReSe2{} has largest optical oscillator strength along the $\mathrm{\Gamma}-\mathrm{K}$ direction for the exciton at $\mathrm{\Gamma}$,  corresponding to a $p$-polarized pump pulse in our geometry as shown in Fig. \ref{fig:1}(b) \cite{Zhong2015,Zhao2015}.  In a single-orientation crystal, the oscillator strength is expected to be suppressed for an $s$-polarised pump pulse, which is oriented along $\mathrm{\Gamma}-\mathrm{M}$ for our geometry.  However,  our SL \ReSe2{} grows in 120$^{\circ}$-rotated domains such that the difference between the three non-equivalent high symmetry points at the BZ corners of the native 1T$^{\prime}$ structure is lost and a peak in oscillator strength occurs along each $\mathrm{\Gamma}-\mathrm{K}$ direction, as shown in Fig. \ref{fig:1}(b) via pink, blue and green ovals.

\begin{figure}[t!] 
\begin{center}
\includegraphics[width=0.49\textwidth]{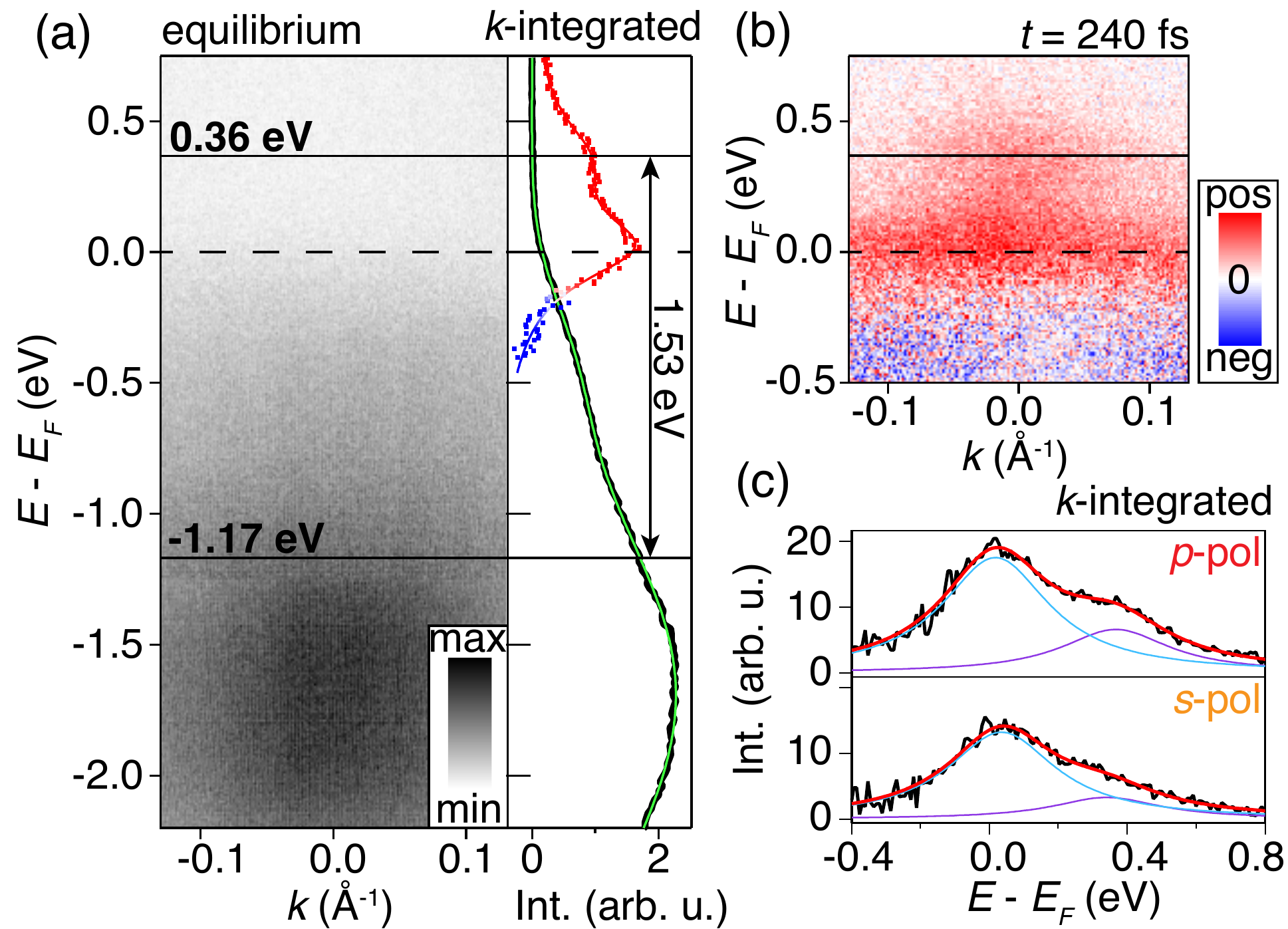}
\caption{(a) Left panel: $(E,k)$-resolved VB spectrum of SL \ReSe2{} acquired in equilibrium conditions using a 10.8 eV laser.  Right panel: $k$-integrated EDC (black curve) of the spectrum with fit (green curve) to a double-exponential rise function. (b) Intensity difference between equilibrium and excited state signal at $t=240$ fs measured using a 1.55 eV $p$-polarised pump pulse and a 6.2 eV probe pulse. The $k$-integrated EDC of the intensity difference is shown in the right panel in (a) with corresponding colorscale.  Fitted peak positions and VB offset are indicated by horizontal lines in (a)-(b) and the values are stated in (a). The resulting optical gap is demarcated by a double-headed arrow in (a). (c) EDCs integrated over $k$ at the time-delay of the peak excitation for measurements with $p$- or $s$-polarised pump pulse.  The EDCs (black curves) are fitted (red curves) by two Lorentzians on a linear background with the individual components shown as blue and purple curves. The linear background of the fit has been subtracted.  The same fitting function has been applied to the intensity difference in the right panel in (a). The $k$-integration range for all EDCs is from -0.05 to 0.05 \AA$^{-1}$.}
\label{fig:2}
\end{center}
\end{figure}

An equilibrium ARPES spectrum of SL \ReSe2{} around the VB maximum at $\mathrm{\Gamma}$,  acquired with the 10.8~eV probe pulse, is shown in Fig. \ref{fig:2}(a).  The left panel presents the $(E,k)$-dependent ARPES intensity while the right panel shows the corresponding energy distribution curve (EDC) integrated over $k$ from -0.05 to 0.05 \AA$^{-1}$.  From the Fermi level $E_F$ (see dashed horizontal line) and towards lower energies the intensity of the background increases monotonically until a broad peak of intensity is reached.  The onset of this broad peak is determined by fitting the $k$-integrated EDC with a double exponential rise function, allowing to discriminate between the rise in background intensity and the rise in intensity caused by the \ReSe2{} VB.  The VB offset determined via this method is $(-1.17 \pm 0.02)$ eV, which is in good agreement with the value of -1.1~eV obtained for a similar sample measured in a synchrotron-based ARPES experiment  \cite{Choi2020}.  The broad and relatively featureless VB dispersion within the narrow $(E,k)$-window we can access with a 10.8 eV probe pulse is consistent with previous experiments and emerges from a manifold of states with a high photoemission cross section in our measurement geometry \cite{Choi2020}.  The background intensity is attributed to in-gap states, which have been observed in STS measurements on SL \ReSe2{} supported on bilayer graphene, and are caused by a distribution of trapped charge impurities at the van der Waals interface between ReSe$_2$ and bilayer graphene \cite{Trinh:2022}.  

The response of  SL \ReSe2{} to optical excitation with a 1.55~eV $p$-polarised pump pulse is presented in Fig. \ref{fig:2}(b) via the $(E,k)$-dependent intensity difference determined by subtracting a spectrum acquired before optical excitation ($t<0$) from a spectrum measured around the peak of the optical excitation at $t=240$ fs.  The excited state is probed using a photon energy of 6.2 eV,  precluding access to excited holes in the VB due to the low photoelectron kinetic energies that can be reached.  An increase of photoemission intensity due to excitation (red contrast in Fig \ref{fig:2}(b)) is observed across the full $k$-range around $E_F$ and in a region centred at higher energy and localized in $k$ around $\mathrm{\Gamma}$.  Extracting an EDC of the intensity difference integrated over $k$ from -0.05 to 0.05 \AA$^{-1}$ and fitting with two Lorentzian peaks on a linear background reveals the lower excitation is centred at (0.02 $\pm$ 0.01) ~eV and the higher at (0.36 $\pm$ 0.01)~eV, as shown in Fig. \ref{fig:2}(a).  The lower excitation is consistent with a charge impurity-induced in-gap state pinned at $E_F$ due to the lack of dispersion and complete delocalization in $k$ \cite{Ulstrup2016,Majchrzak:2021}.  

A gap of $(1.53 \pm 0.02)$ eV is determined between the VB offset and the higher excitation, as seen via double-headed arrows in Fig. \ref{fig:2}(a).  This value is substantially lower than the quasiparticle bandgap of 1.7~eV measured by STS on SL \ReSe2{} on bilayer graphene \cite{Choi2020, Trinh:2022}.  Strikingly, the energy coincides with the lower exciton line in SL \ReSe2{} supported on insulating substrates measured by photoluminescence \cite{Arora2017,Kipczak:2020}, leading to the conclusion that we observe the optical gap and that the exciton is resonantly excited by the infrared pump pulse in our experiment.  The exciton binding energy and quasiparticle gap may vary depending on how the Coulomb interaction is screened by the substrate and by excited charge carriers but the optical gap remains fixed \cite{Qiu2019}.

Changing the pump pulse polarisation from $p$ to $s$ causes a notable decrease of intensity from the exciton state relative to the in-gap state as shown via the $k$-integrated EDCs around the peak of excitation obtained with the two polarisations in Fig. \ref{fig:2}(c). The spectral weight is extracted as the area of Lorentzian profiles fitted to the EDCs (see blue and purple curves in Fig. \ref{fig:2}(c)), revealing a 34 \% reduction of the exciton intensity relative to the in-gap state intensity in the case of $s$-polarisation.  This reduction is consistent with the expected anisotropy of the $\mathrm{\Gamma}$ exciton \cite{Zhong2015}, as indicated by the pink ovals in Fig. \ref{fig:1}(b). 

\begin{figure}[t!] 
\begin{center}
\includegraphics[width=0.49\textwidth]{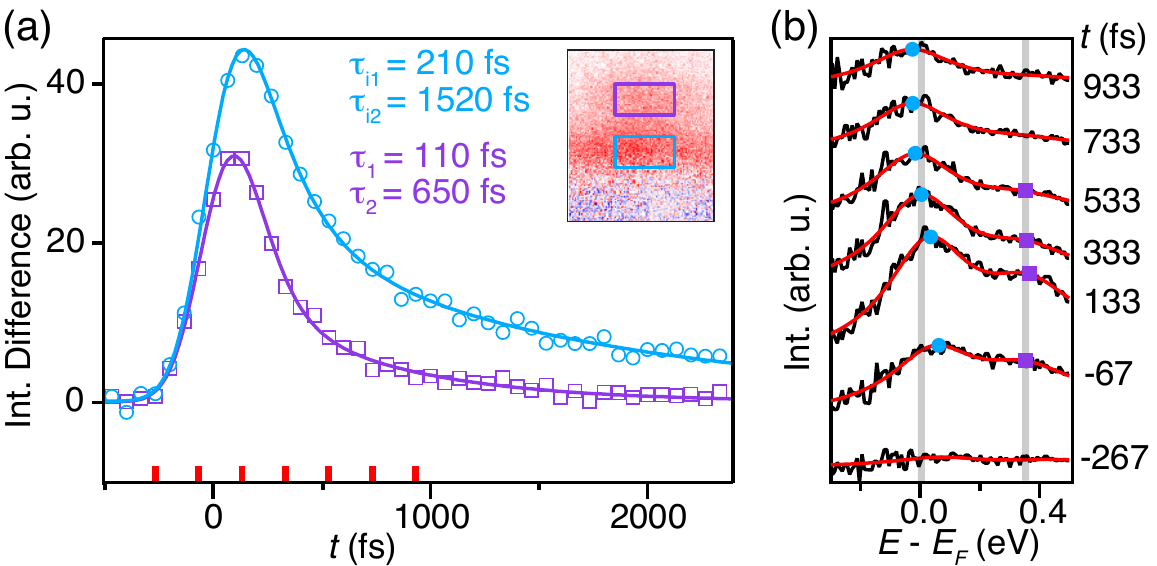}
\caption{(a) Time dependence of the $(E,k)$-integrated intensity difference (circular and square data points) within the correspondingly colored boxes in the inset, which shows the intensity difference from Fig. \ref{fig:2}(b).  Each trace of the intensity as a function of time is fitted with a Gaussian-broadened step function multiplied by a biexponential decay (smooth curves). The decay time constants $\tau_1$, $\tau_{\mathrm{i}1}$,  $\tau_{2}$ and $\tau_{\mathrm{i}2}$ of each fit are stated with corresponding color as the fitted trace.  (b) EDCs (black curves) integrated over $k$ from -0.05 to 0.05 \AA$^{-1}$ for the given time delays (see red tick marks in (a) for indication of time delays). Two Lorentzians on a linear background are fitted (red curves) with peak positions highlighted by blue circles and purple squares.  Vertical grey bars indicate the Fermi energy and the exciton state.}
\label{fig:3}
\end{center}
\end{figure}

The temporal evolution of the exciton and in-gap state is extracted by integrating the intensity difference over corresponding $(E,k)$-regions, as shown in the inset in Fig. \ref{fig:3}(a).  The resulting traces  are shown as purple squares and blue circles in Fig. \ref{fig:3}(a), respectively,  together with fits to a biexponential decay multiplied by a step function and convoluted with a Gaussian that accounts for the time-resolution.  These fits provide time constants for the decay of the exciton state given by $\tau_1 = (110 \pm 10)$ fs and $\tau_2 = (650 \pm 70)$ fs and for the in-gap state given by $\tau_{\mathrm{i}1} = (210 \pm 10)$ fs and $\tau_{\mathrm{i}2} = (1520 \pm 110)$ fs.  The timescales of the exciton signal are significantly shorter than 10 ps and 80 ps reported for SL ReSe$_2$ on an insulating substrate and bulk ReSe$_2$ \cite{Liu:2019,He:18}.  Here, the similar decay constants for the exciton and in-gap states suggests a strong interplay between ultrafast exciton decay and electron-hole recombination via the in-gap state, possibly also involving the bilayer graphene substrate where similar dynamics may be expected \cite{Ulstrup:2014}.  The slower $\tau_{\mathrm{i}2}$ compared to $\tau_2$ is indicative of re-filling of the in-gap state via decay of the exciton,  underscoring the important role of impurity-induced in-gap states for exciton dynamics \cite{Wang:2015a,Wang:2015b}.

We inspect how the EDC peak positions of the signals evolve with time in Fig.  \ref{fig:3}(b).  The exciton peak remains fixed for all time-delays where we could reliably fit a Lorentzian function, which further supports our assignment of this feature in terms of the optical gap and not the quasiparticle gap. The latter would be expected to renormalize as a function of time due to the decaying photoinduced carrier density \cite{chernikov2015population,Ulstrup2016}, thereby leading to a time-dependent peak position. The in-gap state exhibits a peak shift towards $E_F$ due to the decrease in Fermi level broadening as the excited carriers cool down. 

The momentum distribution of the exciton signal measured by TR-ARPES can be related to the spatial distribution of the exciton wavefunction \cite{Shuo2021, Man2021}.  It is thereby possible to determine how localized or delocalized the exciton is in our SL \ReSe2{}/bilayer graphene heterostructure. The photoemission intensity is proportional to the square of the transition matrix element given by $\mathcal{M}_{f,i} = \braket{\psi_{f}| \vec{A} \cdot \vec{p}|\psi_{i}}$, where $\vec{A}$ is the vector potential, $\vec{p}$ is the momentum operator and $\psi_{f}$ ($\psi_{i}$) is the final (initial) state of the photoelectron. The final state can be approximated as a plane wave such that the matrix element becomes proportional to the Fourier transform of the initial state wavefunction, $\mathcal{M}_{f,i} \propto {\vec{A}} \cdot \vec{k}\braket{\vec{k}|\psi_{i}}$,  where $\vec{k}$ is the wave vector and $\bra{\vec{k}}$ signifies a plane wave final state \cite{moser:2017,Shuo2021}. 

\begin{figure}[t!] 
\begin{center}
\includegraphics[width=0.49\textwidth]{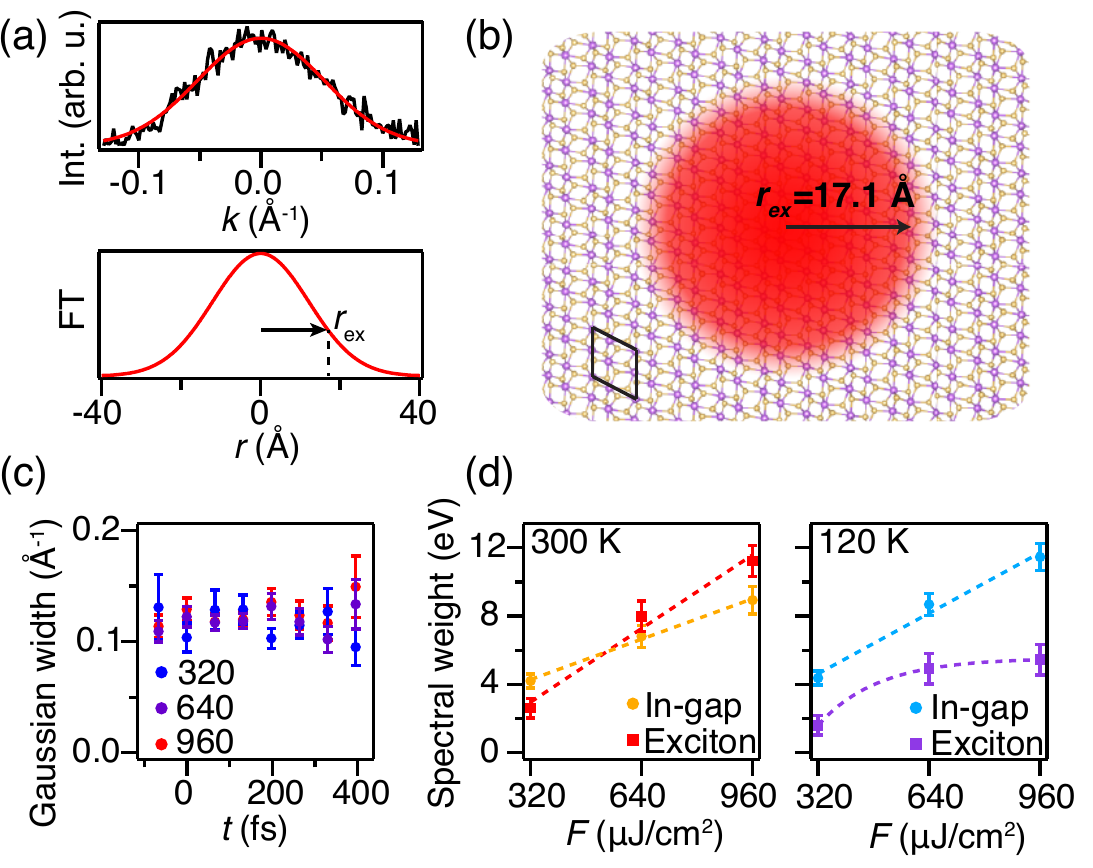}
\caption{(a) Top panel: MDC of the photoemission intensity (black curve) integrated over an energy range from 0.19 to 0.53 eV at the peak of the optical excitation.  A fit to a $k$-dependent Gaussian distribution function is overlaid (red curve).  Bottom panel: Fourier transform (FT) of the Gaussian distribution. The estimate of exciton radius $r_{\mathrm{ex}}$ is indicated by an arrow and a dashed vertical line.  (b) Sketch of the real space distribution of the exciton (red disk) superimposed on the SL \ReSe2{} crystal structure. The unit cell is demarcated by a black rhombus.  (c) Temporal evolution of the width of the $k$-dependent Gaussian distribution for optical excitations with the stated fluence in units of $\mu$J/cm$^2$.  (d) Fluence and temperature dependence of the spectral weight associated with in-gap and exciton signals at the peak of optical excitation.  The spectral weight is extracted as the area of the corresponding Lorentzian fit to the $k$-integrated EDC. Measurements were performed at 300 K (left panel) and 120 K (right panel). Dashed lines are guides to the eye.}
\label{fig:4}
\end{center}
\end{figure}

We utilize this relation by integrating the momentum distribution curve (MDC) of the photoemission intensity over an energy range from 0.19 eV to 0.53 eV, which contains the exciton signal.  Fitting the MDC to a Gaussian function with a width of 0.07 \AA$^{-1}$ provides the momentum distribution of the exciton. This is Fourier transformed to yield the real-space distribution with a half-width at half maximum (radius) given by $r_{\mathrm{ex}} = 17.1$ \AA{}, as shown in Fig. \ref{fig:4}(a). This should be considered a lower estimate due to momentum-broadening factors such as the finite experimental resolution.  The size of the exciton is compared with the SL \ReSe2{} lattice in Fig. \ref{fig:4}(b) and is substantially larger than  9.6 \AA{} determined for bulk \ReSe2{} \cite{Arora2017}.  External screening caused by the bilayer graphene substrate leads to delocalization of the exciton in real space and thereby the larger radius for our SL \ReSe2{} \cite{Stier2016}. Additionally, by measuring the MDC width for different exciton densities, achieved by varying the pump fluence and time delay, we find the width is insensitive to the density within the error bars of our analysis, as shown in Fig. \ref{fig:4}(c). The external screening thereby fully specifies the size of the exciton in our sample.

The spectral weight of the exciton and in-gap states provides a direct measure of their density following optical excitation with different fluence.  In Fig.  \ref{fig:4}(d) we have extracted the spectral weight as the area of Lorentzian fits to EDCs at the peak of excitation, following the approach presented in Fig.  \ref{fig:2}(c). The analysis is performed for three different settings of fluence and at sample temperatures of 120 K and 300 K. The spectral weight increases linearly with fluence for the in-gap states at both temperatures, as expected from the increasing number of photoexcited carriers. In case of the exciton state, the spectral weight saturates with increasing fluence at 120~K but not at 300 K. The exciton absorption strength is known to reach a saturation level due to Pauli exclusion, which limits phase filling and short range screening \cite{Schmitt1985}. At higher temperatures the phase space available to the exciton is greater such that saturation is not reached in our conditions.

In conclusion,  we have applied time- and angle-resolved photoemission spectroscopy to measure exciton population dynamics in SL \ReSe2{} on bilayer graphene. The excited state signal around $\mathrm{\Gamma}$ following infrared optical excitation is dominated by an in-gap state pinned at $E_F$ and an exciton state at (0.36 $\pm$ 0.01)~eV,  leading to an optical gap of $(1.53 \pm 0.02)$ eV.  A reduction of the exciton spectral weight is observed when the polarisation of the optical excitation is changed from $p$ to $s$,  demonstrating a significant optical anisotropy in our sample.  The exciton population exhibits a biexponential decay with time constants of $\tau_1 = (110 \pm 10)$ fs and $\tau_2 = (650 \pm 70)$ fs,  which is facilitated by the in-gap state that displays time constants of $\tau_{\mathrm{i}1} = (210 \pm 10)$ fs and $\tau_{\mathrm{i}2} = (1520 \pm 110)$ fs.  Such drastically improved charge transfer interactions suggest that interlayer interactions make SL \ReSe2{}/graphene heterostructures very attracting for realizing efficient photodetector and field effect transistor applications \cite{Yuan:2018,Park:2021}.  Finally, we estimate the radial extent of the exciton real space distribution to be greater than 17.1 \AA{}, demonstrating significant exciton delocalization due to external screening from the bilayer graphene substrate.  Our observations underpin the significant impact on the optical properties of SL \ReSe2{} from the adjacent bilayer graphene in a van der Waals heterostructure.

We gratefully acknowledge funding from VILLUM FONDEN through the Young Investigator Program (Grant. No. 15375), the Centre of Excellence for Dirac Materials (Grant. No. 11744),  the Danish Council for Independent Research, Natural Sciences under the Sapere Aude program (Grant No. DFF-9064-00057B) and NRF-2019K1A3A7A09033389, 2020R1A2C200373211, 2021R1A6A3A14040322. [Innovative Talent Education Program for Smart City] by MOLIT.

\end{document}